\documentclass[%
 reprint,
nobibnotes,
 amsmath,amssymb,
 aps,
]{revtex4-2}

\usepackage{textalpha}
\usepackage{graphicx}% Include figure files
\usepackage{svg}
\usepackage{dcolumn}% Align table columns on decimal point
\usepackage{bm}% bold math
\usepackage{caption}
\usepackage{subcaption}
\usepackage{color}

\begin{document}

\preprint{APS/123-QED}

\title{Auxetic behavior on demand: a three steps recipe for new designs}% Force line breaks with \\

\author{Daniel Acuna${}^{1,4}$}

 \author{Francisco Gutiérrez${}^{1,4}$}
 
 \author{Rodrigo Silva${}^{2,4}$}
 
 \author{Humberto Palza${}^{2,4}$}
 
\author{Alvaro S. Nunez${}^{1,4,5}$}

\author{Gustavo D\"uring${}^{3,4}$}
\email{gduring@uc.cl}
\affiliation{%
${}^1$Departamento de F\'isica, Facultad de Ciencias Físicas y Matemáticas, Universidad de Chile, Santiago, Chile
}%
\affiliation{%
${}^2$Departamento de Ingeniería Química, Biotecnología y Materiales, Facultad de Ciencias Físicas y Matemáticas, Universidad de Chile, Santiago, Chile
}%
\affiliation{%
${}^3$Instituto de F\'isica, Pontificia Universidad Cat\'olica de Chile, Casilla 306, Santiago, Chile}%
\affiliation{ ${}^4$ANID - Millenium Nucleus of Soft Smart Mechanical Metamaterials, Santiago, Chile}
\affiliation{ ${}^5$CEDENNA, Avda. Ecuador 3493, Santiago, Chile}

\date{\today}% It is always \today, today,

\begin{abstract}

Despite their outstanding mechanical properties, with many industrial applications, a rational and systematic design of new and controlled auxetic materials remains poorly developed. Here a unified framework is established to describe bidimensional perfect auxetics with potential use in the design of new materials. Perfect auxetics are characterized by a Poisson's ratio $\nu=-1$ over a finite strain range and can be modeled as materials composed of rotating rigid units. Inspired by a natural connection between these rotating rigid units with an antiferromagnetic spin system, here are unveiled the conditions for the emergence of a non-trivial floppy mode responsible for the auxetic behavior. Furthermore, this model paves a simple pathway for the design of new auxetic materials, based on three simple steps, which set the sufficient connectivity and geometrical constraints for perfect auxetics. In particular, a new exotic crystal, a Penrose quasi-crystal and the long desired isotropic auxetic material are designed and constructed for the first time. Using 3D printed materials, finite element methods and this rigid unit model, the auxetic behavior of these designs is shown to be robust under small disturbances in the structure, though the Poisson's ratio value relies on system's details, approaching $-1$ close to the ideal case.

\end{abstract}

\maketitle

\section{Introduction}

The design of new materials with unusual mechanical properties and advanced functionalities has become a very active field of research in soft matter physics. These so-called mechanical metamaterials acquire their mysterious behavior from the particular inner architecture and not from the constituent materials properties. In recent years, metamaterials have been engineered to display topological protection\cite{huber2016topological, coulais2017static,kane2014topological, chen2014nonlinear, rocklin2017transformable, souslov2017topological}, programmable shapes  \cite{florijn2014programmable, coulais2016combinatorial, overvelde2017rational, mullin2007pattern, shim2012buckling}, nonlinear response \cite{mullin2007pattern, nicolaou2012mechanical, shim2012buckling, coulais2015discontinuous, wyart2008elasticity, coulais2017static} and negative elastic constants \cite{lakes1987foam, coulais2015discontinuous, bertoldi2010negative, nicolaou2012mechanical, lakes2001extreme} among others. Auxetic materials are probably the epitome of mechanical metamaterials, which were for the first time intentionally designed by Lakes in 1987 \cite{lakes1987foam}.
An auxetic material, unlike common elastic materials, when compressed (expanded) in a given direction, compresses (expands) in the perpendicular direction. This unusual property is characterized by a negative Poisson's ratio $\nu$, the ratio between the strain in one direction and the strain in its perpendicular direction.  

A negative Poisson's ratio has been found in natural bioauxetics \cite{williams1982properties,pagliara2014auxetic} and molecular auxetics \cite{baughman1998negative,grima2005origin}. Nowadays, with the onset of 3D printing, a wide range of auxetic materials are being developed \cite{coulais2018characteristic,courentinDomainWall,bertoldi2010negative,coulais2015discontinuous,shim2012buckling,babaee20133d}, with interest in their enhanced mechanical properties, like increased energy absorption\cite{CHENenergy}, enhanced indentation resistance\cite{LakesIndentation}, high fracture toughness\cite{YANGfracture}, synclastic curvature in bending\cite{lakes2002making}, and variable permeability\cite{AldersonPorus2}, with applications in bio-medicine\cite{gatHierarchical} and textiles\cite{AldersonTextil} as some examples.

\begin{figure*}[t]
\includegraphics[width=\linewidth]{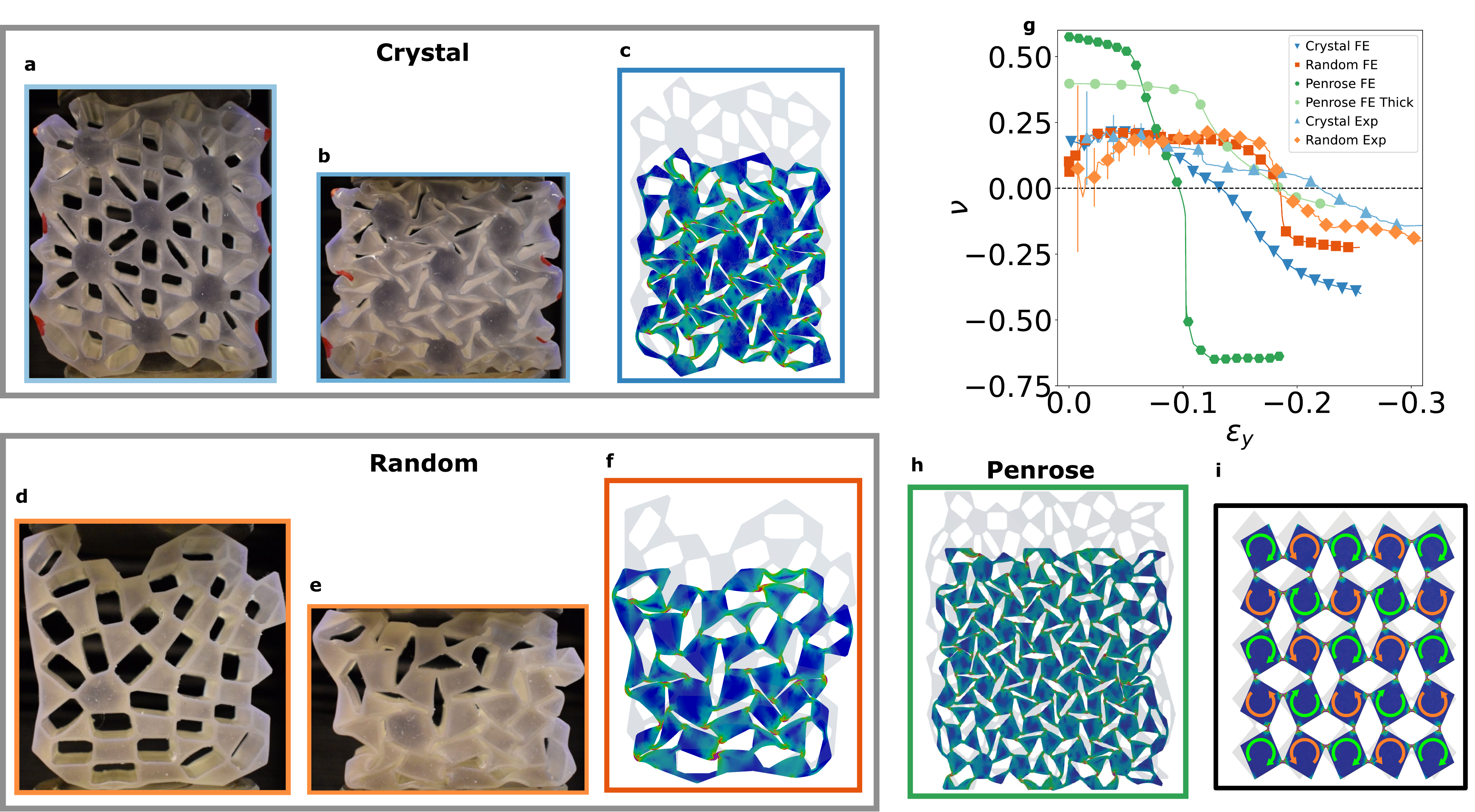}
\caption{\label{fig:redes} \textbf{Auxetic behavior on demand}. Our proposed algorithm generated three instances of auxetic materials. The three systems correspond to: \textbf{a, b, c)} an Exotic Crystal, \textbf{d, e, f)} a Random Lattice (isotropic lattice), and \textbf{h)} a Penrose quasicrystal.  The structures in \textbf{c, f, h)} are finite element simulations, the color shows the intensity of the stress in the material, blue being low stress and red being high stress. The structures in \textbf{a, d)} were 3D printed on elastic resin. Their respective uniaxial compression results are depicted in \textbf{b, e)}(Video S1, Supporting Information) and in their finite element simulations \textbf{c, f, h)}(Video S2, Supporting Information), in both the auxetic behavior is apparent. This unusual property's basic mechanism is the coordination and synchronization of the buckling instability at each of the weak links that provide the structure its stability. The effective collective pattern that emerges is analogue to an antiferromagnetic arrangement. Each of the two interconnected lattices that fit the bipartite system rotates in opposite  directions, as illustrated in \textbf{i)}. Out of this analogy, we infer that several properties of the anisotropic XY antiferromagnet\cite{2006Mattis} are inherited into the context of auxetic systems. In \textbf{g)}, we display the Poisson's ratios of each configuration, ``Exp'' and ``FE'' refer to experimental and finite element simulation results respectively. Noteworthy is the comparison between a Penrose structure with thin (dark green hexagons) and thick (light green circles) bonds between polygons, showing the capabilities of tuning the Poisson's ratio as a function of the bending energy.}
\end{figure*}

A variety of shapes and geometries have been identified as prototypical auxetics. The list ranges from re-entrant structures\cite{reentrant1} to rotating units\cite{grima2000auxetic,alderson2001rotation}. Chiral structures\cite{chiral} and retractile bars systems \cite{rens2019rigidity,MILTON19921105} complete the list. 
Despite the extensive literature and enormous  progress describing different types of auxetic materials no fundamental microscopic principles for a unified description  exist. The distinction between types of auxetics relies mainly on empirical observation rather than in fundamental principles, and no general prescription exists to build them. In this article, we present a unified framework for the description of bi-dimensional perfect auxetics. In this limit the bulk modulus vanishes while the shear modulus remains finite implying a perfect auxetic behavior, i.e. with a Poisson's ratio $\nu=-1$ over a finite strain range. Understanding and controlling the perfect auxetic limit allows us to rationally design realistic auxetics with a tailor made geometric structure for which the Poisson's ratio remains negative. The precise value of the Poisson's ratio depends on the proximity to the ideal design and the energy interaction between the material constituents.

To our knowledge rotating unit models and retractile bars systems are the only ones that have displayed a perfect auxetic behavior, interestingly both materials are quite related \cite{MILTON19921105}. Rotating unit systems \cite{grimaRotTriangles,gatHierarchical,grima2000auxetic,grima2007auxetic,BertoldiDomainWall,courentinDomainWall,alderson2001rotation,bertoldi2010negative} can be considered as a material made of elastic units connected to their neighbors in such a way that the energy cost of deforming the bulk of each unit is much higher than that of deforming the bond between neighbors. Generically one can think of a structure made out of polygons connected through their vertices, \textbf{Figure \ref{fig:redes}}i is a common example composed of squares. Under external loads the stresses focalize on the vertices leaving the bulk of the polygons almost undeformed. The auxetic behavior arises because neighbor polygons tend to rotate in opposite directions along a particular low energy mode. This mode is reminiscent of a non-trivial floppy mode, or mechanism, that exists in the limit case with zero bending energy (i.e. the polygons are connected through ideal hinges) which leads to a perfect auxetic behavior \cite{grima2007auxetic}. If polygons are considered as rigid structures, an extended version of the Maxwell's degrees of freedom counting argument \cite{maxwell1864calculation} shows that auxetic polygon networks of rotating unit systems are isostatic or overconstrained (Appendix \ref{DoF}). Therefore, the existence of an ``auxetic'' floppy mode must be related to a very precise geometrical construction which also implies the appearance of a non-trivial self stress state mode following the rank theorem \cite{calladine1978buckminster}. No general conditions for the emergence of this floppy mode exist, except for certain limited sets of periodic lattices \cite{GUEST2003383,mitschke2013symmetry}. Other auxetics have different origins, for instance, reentrant materials have typically an under-constrained internal structure stabilized by bending or angular forces \cite{reentrantReid}. Therefore, the shear and the bulk modulus vanishes in the zero bending limit, excluding a priori the existence of a perfect auxetic behavior.

To understand the auxetic behavior of rotating units systems we restrict our study to their perfect auxetic limit. This can be easily achieved considering a minimal model of rigid polygons connected by springs of zero natural length \cite{BertoldiDomainWall}. The springs act as ideal hinges as long as they are not compressed or stretched. Since rigid polygons can only rotate, any floppy mode requires that all the neighbors of each polygon have the same rotation rate (as a function of strain). If the neighbors of a given polygon are also neighbors between them the system will then jam. A similar behavior is observed for antiferromagnetic spin systems, if the neighbors of a spin are also neighbors between them the antiferromagnetic phase will be frustrated. A pure antiferromagnetic phase is known to be achieved only for bipartite networks. This similarity sets the key ingredient for rotating units auxetic theory, which requires the system to be bipartite as well, as can be easily check for all previous rotating units auxetics. As we will discuss later, the connection with spin systems can be pushed further by looking at the potential energy of our model which can be mapped to an anisotopic antiferromagnetic spin systems. Exchanging temperature with strain a Ginzburg-Landau energy can be constructed leading to a general descriptions of domain walls in auxetic materials, a phenomena recently observed in \cite{courentinDomainWall,BertoldiDomainWall}. Although bipartite polygon networks is a necessary condition, and most of them show some level of auxetic response, additional conditions are necessary to obtain a perfect auxetic behavior.

Our theory not only unifies a large class of existing auxetics, encompassing previous auxetic rigid units systems, but also allows to create a large variety of them, with new and diverse geometries, including the somewhat elusive isotropic auxetic material \cite{grimaKirigami,reentrantReid}. In Figure \ref{fig:redes} one can see three different examples of elastic structures based on perfect auxetic designs; a new type of auxetic crystals, a quasicrystal and an isotropic (disordered) structure. These structures where 3D printed on elastic resin, their mechanical response was measured and compared against finite element simulations performed on the software Ansys (Appendix \ref{FEM}). After a finite initial load, they display a clear auxetic behavior with a Poisson's ratio reaching values between $-0.62$ and $-0.14$. The precise value of the Poisson's ratio depends mainly on how close the design is to the ideal scenario, i.e. no bending energy. This behavior can be easily controlled by changing the thickness of the bonds between polygons \cite{bertoldi2010negative}. An example of this is shown in Figure \ref{fig:redes}g, where two Penrose auxetics with different bond thickness are compared, the design with thinner bonds and therefore lower bending energy shows a smaller Poisson's ratio, which is much closer to the ideal value. As these structures are made from perfect auxetic designs, decreasing the bonds thickness will result in a lower Poisson's ratio, finally reaching the ideal limit $\nu=-1$ when the bonds vanish and become hinges.

\section{ A simple model for perfect auxetics}

We are interested in the behavior of perfect auxetics, to build them we use  the minimal bi-dimensional model for rotating units auxetics, consisting of a series of polygons connected by springs of zero natural length \cite{BertoldiDomainWall}. Each rigid unit has three degrees of freedom, two translational $\vec{x}_i=\left(x_i,y_i \right)$ and one rotational $\theta_i$, not necessarily measured from the centroid of each polygon. This model allow to determine the emergence of the auxetic  floppy mode and also introduces elasticity into our materials to study non ideal scenarios. The recipe for building a perfect auxetic requires three ingredients.

\begin{figure*}[t]
\centering
\includegraphics[width=0.8\textwidth]{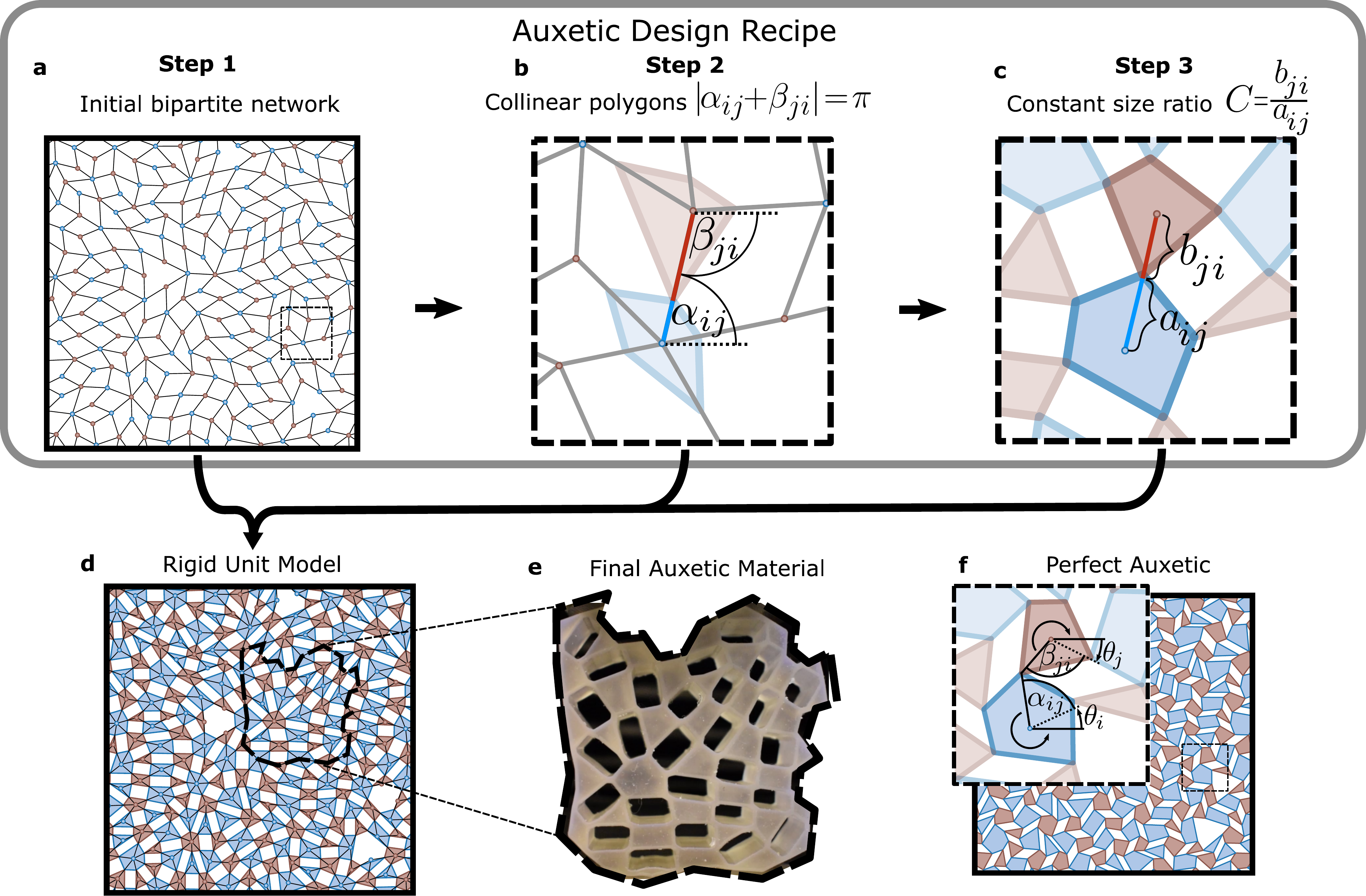}
\caption{\textbf{The 3 steps recipe}. \textbf{a)} Step 1, we start with a bipartite network, the blue and red colors stand for the A and B sets respectively. We showcase a random bipartite network to demonstrate the versatility of this recipe. \textbf{b)} Step 2, every node of the graph becomes the position of each polygon and we place each polygon's vertex on top of each corresponding segment of the network. At rest the polygon's position is collinear with its neighbor's position and the vertex between them. \textbf{c)} Step 3, each polygon vertex is positioned such that the ratio $C=b_{ji}/a_{ij}$ remains constant along the network.  $a_{ij}$ is the distance from the node to each vertex in the A set polygons, $b_{ji}$ is the analogue for the B set. \textbf{d)} The resulting polygon network after performing the 3 steps. \textbf{e)} A section of the final random auxetic design is printed in 3D, as every section of the model is inherently a perfect auxetic, the printed material keeps being auxetic. \textbf{f)} A compression shows that the polygon network is a perfect auxetic. In the zoom in we see how each polygon counter rotates with respect to its neighbors. $\theta_i$ and $\theta_j$ show the rotation of the A and B set respectively.}
\label{fig:net2Aux}
\end{figure*}

\begin{enumerate}
\item {\bf The network must be bipartite. }

This allows the units to counter rotate respect to each other, like cogs in a machine.
The units  arranged in a bipartite network can be separated in two sets  $A$ and $B$, i.e. each connected to the other but not to itself, see \textbf{Figure \ref{fig:net2Aux}}a.

\item {\bf Initially and at rest, every pair of neighboring polygons position's ($\vec{x}_i$, $\vec{x}_j$) and the vertex between them have to be collinear. }

This initial setting, matches a maximum extension configuration. Furthermore, it establishes a relationship between the internal angles of every pair of neighboring polygons $|\alpha_{ij}+\beta_{ji}|=\pi$, see Figure \ref{fig:net2Aux}b.

\item {\bf The ratio between the distance of a polygon to one of its vertex and the distance of his neighbor to the same vertex must be a constant in the network. }

Each vertex of a rigid unit is characterized  by a vector $\vec{a}_{ij}=a_{ij}\left(\cos(\theta_i+\alpha_{ij}),\sin(\theta_i+\alpha_{ij}) \right)$ or  $\vec{b}_{ji}=b_{ji}\left(\cos(-\theta_j-\beta_{ji}),\sin(-\theta_j-\beta_{ji}) \right)$, corresponding to sets $A$ or $B$ respectively. The index $i$ will be used for polygons in the set $A$ and the index $j$ for polygons in the set $B$. Vectors  $\vec{a}_{ij}$ ($\vec{b}_{ji}$) point from the position $\vec{x}$ of the polygon $i(j)$ into the vertex connecting with polygon $j(i)$, as seen in Figure \ref{fig:net2Aux}c and \ref{fig:net2Aux}f. Therefore, the ratio $C=b_{ji}/a_{ij}$ must be a constant through the network.
\end{enumerate}

Creating a polygon network that fulfills these rules is quite simple. Starting from a planar bipartite graph one can always build a perfect auxetic. To understand the origin of this behavior we turn to the energy of the polygon network

\begin{equation}
\label{eq:springLam}
V=\frac{k}{2}\sum_{<ij>}\left( (\vec{x}_i+\vec{a}_{ij})  -(\vec{x}_j+\vec{b}_{ji})  \right)^2,
\end{equation}
where the sum is over all the pairs of interacting neighbors and all springs have an equal elastic coefficient $k$.

We shall consider the energy change under an isotropic compression, which is equivalent to increasing the size of all polygons while keeping the distance between polygons constant. From the second requirement, as the neighboring polygons are initially collinear with their vertex, the the constant distance between polygons is $\vec{x}_i-\vec{x}_j=-(a_{ij}+b_{ji}) \left( \begin{array}{c} \cos(\alpha_{ij})\\ \sin(\alpha_{ij}) \end{array} \right)$. Increasing the size of the polygons is achieved by rescaling with $\lambda$ the vectors characterizing the vertices, such that $\vec{a}_{ij}\rightarrow \lambda \vec{a}_{ij}$ and $\vec{b}_{ji}\rightarrow \lambda \vec{b}_{ji}$. Then we can expand and rearrange equation (\ref{eq:springLam}) as

\begin{equation}
\label{eq:ising}
V=V_0+\sum_{<ij>}J_{ij}\cos(\theta_i+\theta_j)-H_{ij}^A\cos(\theta_i)-H_{ji}^B\cos(\theta_j).
\end{equation}
In this equation $V_0=\frac{k}{2}\sum_{<ij>} (a_{ij}+b_{ji})^2+\lambda^2(a_{ij}^2+b_{ji}^2)$, $J_{ij}=k \lambda^2 a_{ij}b_{ji}$, $H_{ij}^A=k\lambda (a_{ij}+b_{ji})a_{ij}$ and $H_{ji}^B=k\lambda (b_{ji}+a_{ij})b_{ji}$.

Using the third requirement, setting $C=b_{ji}/a_{ij}$ as a constant, two solutions can be found. The first one is the trivial solution $\theta_i=\theta_j=0$ which is a minimum for $0<\lambda\leq 1$. The second one is found when all the polygons of each set rotate at the same rate $\theta_i=\theta^0_A$ and $\theta_j=\theta^0_B$, where

\begin{equation}
\label{eq:minima1}
\cos(\theta^0_A)=\frac{1+C+\lambda^2(1-C)}{2\lambda},
\end{equation}

and

\begin{equation}
\label{eq:minima2}
\cos(\theta^0_B)=\frac{1+C-\lambda^2(1-C)}{2\lambda C}.
\end{equation}

 These are a minimum in the range $1<\lambda<\big|\frac{1+C}{1-C}\big|$ only if both $\theta^0_A$ and $\theta^0_B$ have the same sign, i.e. polygons counter rotate respect to each other. Evaluating the potential energy in this minimum we find that $V(\theta_i=\theta^0_A,\theta_j=\theta^0_B)=0$ Appendix \ref{PerfectAuxeticDerivation}), thus this solution describes a zero energy mode of the system. This floppy mode corresponds to a system with zero bulk modulus, meaning that the material expands and contracts equally in all directions, for a direct calculation of the bulk modulus see Appendix \ref{bulk}. As the Poisson's ratio is defined as $\nu=-\frac{d\epsilon_x}{d\epsilon_y}$, with $\epsilon$ being the strain in each direction, if the system expands equally in both directions then $\epsilon_x=\epsilon_y$ and $\nu=-1$. This perfect auxetic behavior can be seen in the numerical simulations of periodic and isotropic materials, in \textbf{Figure \ref{fig:frustrations}}d and \ref{fig:frustrations}e.

\section{Random perfect auxetics}

Recently, several isotropic auxetics materials with a Poisson's ratio close to $-1$ in an infinitesimal strain range have been proposed \cite{grimaKirigami,reentrantReid}. However, none of them display a perfect auxetic behavior. The theory presented previously is not restricted to lattices. It can be applied straightforwardly to disordered networks, which allow us to build for the first time isotropic perfect auxetic materials. The only complication resides in building a planar bipartite disordered network. We achieved this by two different methods, a pruning algorithm, and a graph transformation, both applied on an isotropic amorphous contact network \cite{lerner2014breakdown,wyart2008elasticity} (Appendix \ref{builBipartite}). Once we have our bipartite network, we only need to place polygons at each node of the graph, taking care that each vertex of the polygon is placed over a segment of the graph, dividing it in a $C:1$ ratio, see Figure \ref{fig:net2Aux}.

\section{New Section}

Strikingly, the structure of equation (\ref{eq:ising}) and the counter-rotation of the neighboring angles resembles that of an anisotropic antiferromagnetic spin system. Moreover the system has a critical point $\lambda=1$ where the polygon's angle goes from zero to a rotated state, equation (\ref{eq:minima1}) and (\ref{eq:minima2}). Considering these factors and neglecting the polygon displacements we can build an energy functional for this model, with a normalized polygon angle $\theta(x)$ as a order parameter and the expansion factor $\delta \lambda=\lambda-1$ as a control parameter,

\begin{equation}
\label{eq:energyF}
H=\int dx \left( \frac{-t_0\delta \lambda}{2}\theta(x)^2+u\theta(x)^4+\frac{K}{2}\left(\nabla\theta(x)\right)^2 \right).
\end{equation}

\begin{figure*}[t]
\centering
\includegraphics[width=\linewidth]{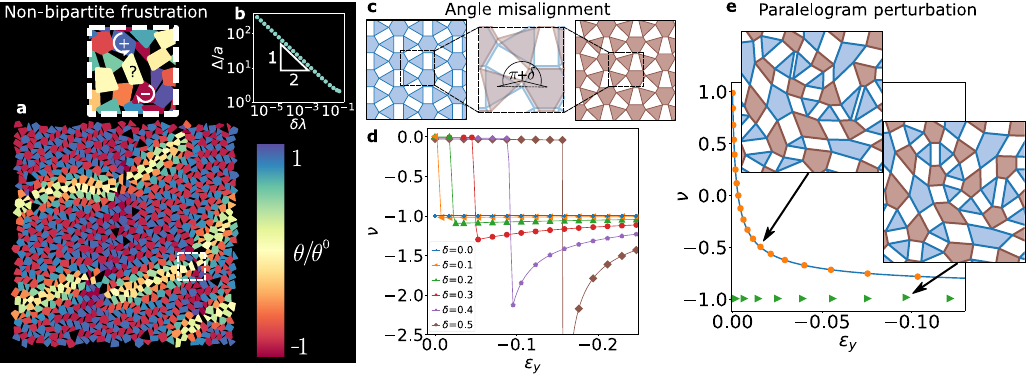}
\caption{\label{fig:frustrations}  \textbf{Breaking the rules.} Three interesting examples that break the rules to build a perfect auxetic. \textbf{a)} By breaking the bipartivity the auxetic will frustrate, creating local defects that don't rotate as predicted. Here we showcase a non-bipartite random polygon network, when compressed a system spanning defect appears (yellow), reminiscent of a domain wall. The color represents the rotation of each polygon, red being clockwise and blue counter clock wise, yellow shows the frustrated polygons. To create this defect, periodic boundary conditions where applied to an initially bipartite network, this boundary condition created a non-local odd cycle which destroyed the bipartivity. \textbf{b)} The domain wall length $\Delta$, normalized by the polygon size $a$, plotted against the isotropic expansion parameter $\delta \lambda$, it's evident that $\Delta\sim\delta\lambda^{-1/2}$, more on Appendix \ref{DomainWalls}. \textbf{c)} Misaligning pairs of polygons by an angle $\delta$, leaves them no longer collinear between them and their vertex. Notice that this perturbation turns the previously parallelogram holes into trapezoids. This effectively jams the system, delaying its auxetic response as seen in the Poisson's ratio in \textbf{d)} (Video 3, Supporting Information). \textbf{e)} Perturbating the geometry of a perfect auxetic (green triangles) while preserving the parallelogram property of the holes, results in the preservation of the non-trivial floppy mode in the system, while changing it's Poisson's ratio (orange circles). Surprisingly the new Poisson's ratio matches perfectly with that of a rectangular biholar auxetic (Video S4, Supporting Information). For more information about the measurements, see Appendix \ref{numerical}.
 }
\end{figure*}

Where $t_0$, $u$ and $K$ are positive constants, and from expanding equation (\ref{eq:minima1}) and (\ref{eq:minima2}) $u=\frac{C}{8}t_0$. This model predicts the appearance domain walls like the one in Figure \ref{fig:frustrations}a, with a characteristic length $\Delta\sim \delta \lambda^{-1/2}$, which perfectly agrees with the numerical simulations of periodic perfect auxetics, Figure \ref{fig:frustrations}b, the exact calculations and simulations are shown in Appendix \ref{DomainWalls}.

\section{Beyond perfect auxetics}

The bipartite condition seems to be fundamental to have an auxetic material composed of rotating units. Introducing defects on the network will frustrate the rotation of the units affecting the auxetic behavior, Figure \ref{fig:frustrations}a. We consider the effect of breaking other conditions, for example by modifying the angle that connects two polygons, thus making $|\alpha_{ij}+\beta_{ji}|=\pi+\delta_{ij}$. For the sake of simplicity we used a periodic polygon network and a fixed $\delta$ that will change sign from one link to the next, see Figure \ref{fig:frustrations}c. 
By perturbing the polygon network the floppy mode is destroyed and the polygon network jams. Under compression the system now increases its energy by keeping initially a Poisson's ratio close to zero (see Figure \ref{fig:frustrations}d). However at a finite strain an elastic instability occurs and the system jumps to a rotated configuration recovering its auxetic behavior.% as seen in Figure \ref{fig:frustrations}. %The required strain for this transition scales with $\delta^2$, this can be checked by introducing $\delta$ into Eq.~\ref{eq:ising} {\blue Falta agregar appendice}.

Although the latter behavior seem to be generic under small perturbations of the perfect auxetic network it is not always the case. It is known that rectangular networks \cite{courentinDomainWall}, which are not perfect auxetics, preserve a floppy mode and the Poisson's ratio moves continuously from positive to negative values. To understand the condition for the existence of this floppy mode we need a more general description on polygon bipartite networks.

\section{Floppy modes on bipartite polygon networks}

Bipartite graphs have only even sided cycles, these are closed paths that start and end at the same node. The simplest of cycles are the faces of the graph, which are the regions bounded by edges. When transforming a bipartite graph into a polygon network, even cycles are reflected at the geometry of empty spaces between the polygons which are also enclosed by the same number of sides. We will refer to this empty spaces as holes, and to the vertex between polygons as hinges. Notice that no odd sided holes can exist in bipartite systems. In the case of our perfect auxetic materials, one can use the Varignon's quadrilateral theorem\cite{coxeter_greitzer_1967} and Thales theorem to show that the constant ratio $C=b_{ji}/a_{ij}$ directly implies that all 4 sided holes will be parallelograms. 

For a system to be deformed while at zero energy, it needs the opposite angles at each hinge to have an opposite deformation rate. This condition is extremely difficult to fulfill if the inner angles inside a hole have a non-linear behavior as a function of the deformation, as is the case with trapezoidal 4 sided holes. Now parallelograms have the special property that when deformed, all of their inner angles share the same deformation rate, except for the sign. All 4 sided holes in perfect auxetics are parallelograms, this allows every inner angle in each hole to be linearly related to the deformation, fulfilling the restriction at each hinge. This mechanism suggests that any bipartite polygon network with only parallelogram holes will have a non-trivial floppy mode, however, it will not necessarily behave like a perfect auxetic. This inner angles analysis is similar to the ``vertex model''\cite{bossart2020oligomodal}.

In particular, when geometrically perturbing the polygons of a perfect auxetic with only 4 sided holes, while preserving their parallelogram shape, the floppy mode persists. The simplest example is the rectangular network studied in \cite{courentinDomainWall,grima2000auxetic} where the Poisson's ratio was observed to change its sign depending on the strain. Similar results are observed in a more general case where 4 sided holes remain parallelograms after a perturbation of an isotropic perfect auxetic, showing a continuous change from positive to negative Poisson's ratios under strain as seen in Figure \ref{fig:frustrations}e.

\section{Conclusion}

We have presented a simple model that builds the necessary framework to create, design, and characterize rotating unit auxetics. Such framework is built upon a simple analogy between the rotating unit auxetics and an anisotropic XY antiferromagnetic system. As shown in Figure \ref{fig:redes}, we have applied those ideas to generate novel auxetic structures in the form of a crystal, a quasicrystal, and a random lattice.
Each design can be represented, within our theory, by a minimal model, based upon polygons and springs that captures its essential collective response to external loads. 
These models can be simulated straightforwardly to test materials properties while ignoring bending forces. However, if needed, bending can easily be added to the model.
As we have seen, this model correctly describes the behavior of all rotating units systems and could be used to predict new behaviors.
In particular, we have generalized the behavior of auxetic domain walls, which are natural textures that these systems have because of the analogy with magnetic systems.
More phenomena related to this analogy remain to be seen and encourage further investigation.
As a major tangible result, our work leads us to establish the ground rules to create never seen before isotropic perfect auxetics.
This model still lacks dynamic and vibrational analysis that, hopefully, will shed light on the general topological properties of these materials.
Furthermore, this theory has the potential to be expanded into 3D polyhedral materials, as it is mainly written in a vectorial form, we are excited to see where this takes us.

\section{Experimental Section}

\textit{Network Design}: For the simulations and experiments in Figure \ref{fig:redes}, we used three bipartite networks which were created using the methods described in Appendix \ref{builBipartite}. For the exotic crystal we modified a tetrakis tiling by skewing it and applying the bipartite transformation; for the quasi-crystal we used a Penrose tiling which was cut in a suitable square shape; lastly the random network was created using the pruning method.  All the materials were built with the same ratio between average polygon size and bond thickness, such that they exhibit a similar behavior.

\textit{Materials}: A commercial elastic resin (code name Elastic $50A$) from Formlabs was used for the direct 3D printing of the auxetic structures. From the several available resins, Elastic $50A$ presented the elasticity needed for the tests further allowing a printed structure with the proper resolution. The mechanical characteristics of the material are $160\%$ of elongation at failure, $19.1~ kN ~m^{-1}$ of tear strength, $3.23~ MPa$ of ultimate tensile strength and $50A$ of shore hardness \cite{Elastano}.

\textit{3D Printing}: A Form 2 3D printer from Formlabs was used to print directly the auxetic structures avoiding the need to print a negative mold. The model file was loaded in the Preform software of the 3D printer for size adjustments and the addition of the corresponding supports. We tested different sizes until the direct printing allowed a structure with the proper elasticity and resolution avoiding defects and imperfections. The structures were defined having at least $40$ rigid units. The final dimensions of the structures (height $\times$ width $\times$ thickness) were $56~\times~57~\times~ 20~ mm$ for the random structure and $68~ \times~55~\times~ 20~ mm$ for the exotic crystal, they were printed with $100~ \mu m$ of layer thickness resolution.

The compression tests were carried out in a Jinan mechanical tester model WDW-S5 using a $5~ kN$ cell and a velocity of $0.83~ mm~ s^{-1}$. The compression was applied until a $30\%$ of deformation approximately.  A professional camera Nikon D5300 was used to record the sample during compression and expansion. From these images we tracked the position of each rigid unit, which we used to measure the Poisson's ratio.

\textit{Finite Element Simulations}: For our static finite elements simulations, we used the commercial software ANSYS \cite{Ansys} and a Neo-Hookean energy density as a material model, with an initial shear modulus, $G=0.15~ MPa$, and Poisson's ratio $\nu=0.5$, and in-plane strain conditions with hybrid quadratic triangular elements (ANSYS type PLANE183 \cite{AnsysAPDL}). 
We carried out a mesh optimization to ensure that bonds, where most of the strain and stress is localized, are meshed with at least three elements. 
This way, the material has $10^5$ elements approximately.
To compress the material uniaxially, we applied a vertical displacement of the top row of polygons, and fixed the position of the bottom row of polygons. We imposed a free boundary condition in the horizontal direction and we imposed a no-penetration condition, therefore the material can contact itself.

\section{Acknowledgments}

A.S.N. thanks Fondecyt Regular 1190324 and Financiamiento Basal para Centros Cient\'ificos  y  Tecnol\'ogicos  de  Excelencia FB 0807.
D.A. acknowledges funding by the National Agency for Research and Development (ANID) /  Scholarship Program / DOCTORADO NACIONAL/2019 - 21192070. 
This work was funded by ANID - Millenium Science Initiative Program - Code NCN17\_092.
We also thank C. Coulais and C. Falcon for discussions.

\bibliography{biblio.bib}% Produces the bibliography via BibTeX.

\clearpage

\onecolumngrid
\appendix

\section{Building Random Bipartite Planar Graphs}
\label{builBipartite}

\begin{figure}
\centering
\includegraphics[width=1\linewidth]{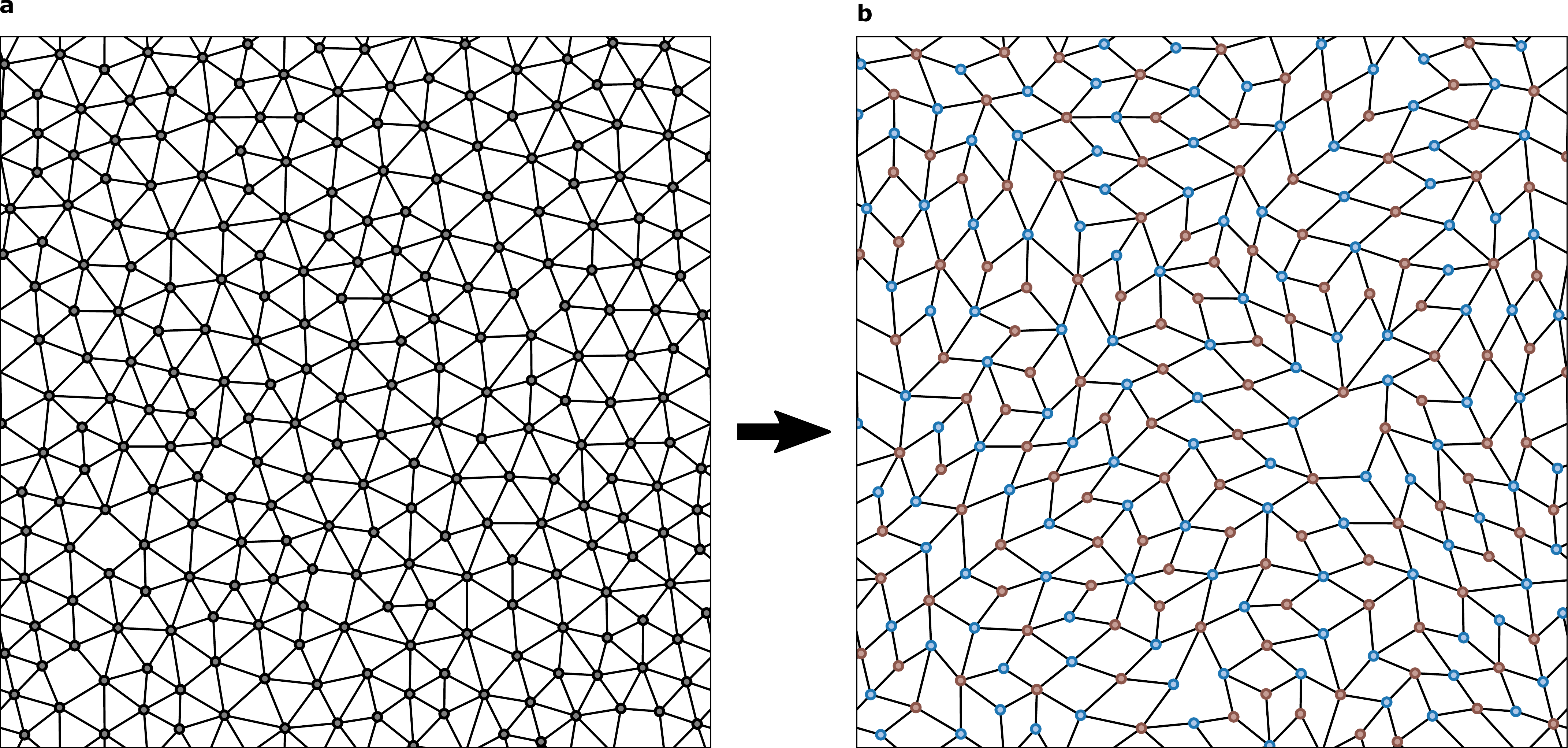}
\caption{\label{fig:prune} \textbf{Pruning Algorithm.} \textbf{a)} Isotropic contact amorphous network \cite{lerner2014breakdown,wyart2008elasticity} with a high coordination $z=6$. \textbf{b)} Pruned network, the bonds between pairs of odd cycles have been removed, adding them together into even cycles. The result is a bipartite planar graph.
 }
\end{figure}

\begin{figure}
\centering
\includegraphics[width=1\linewidth]{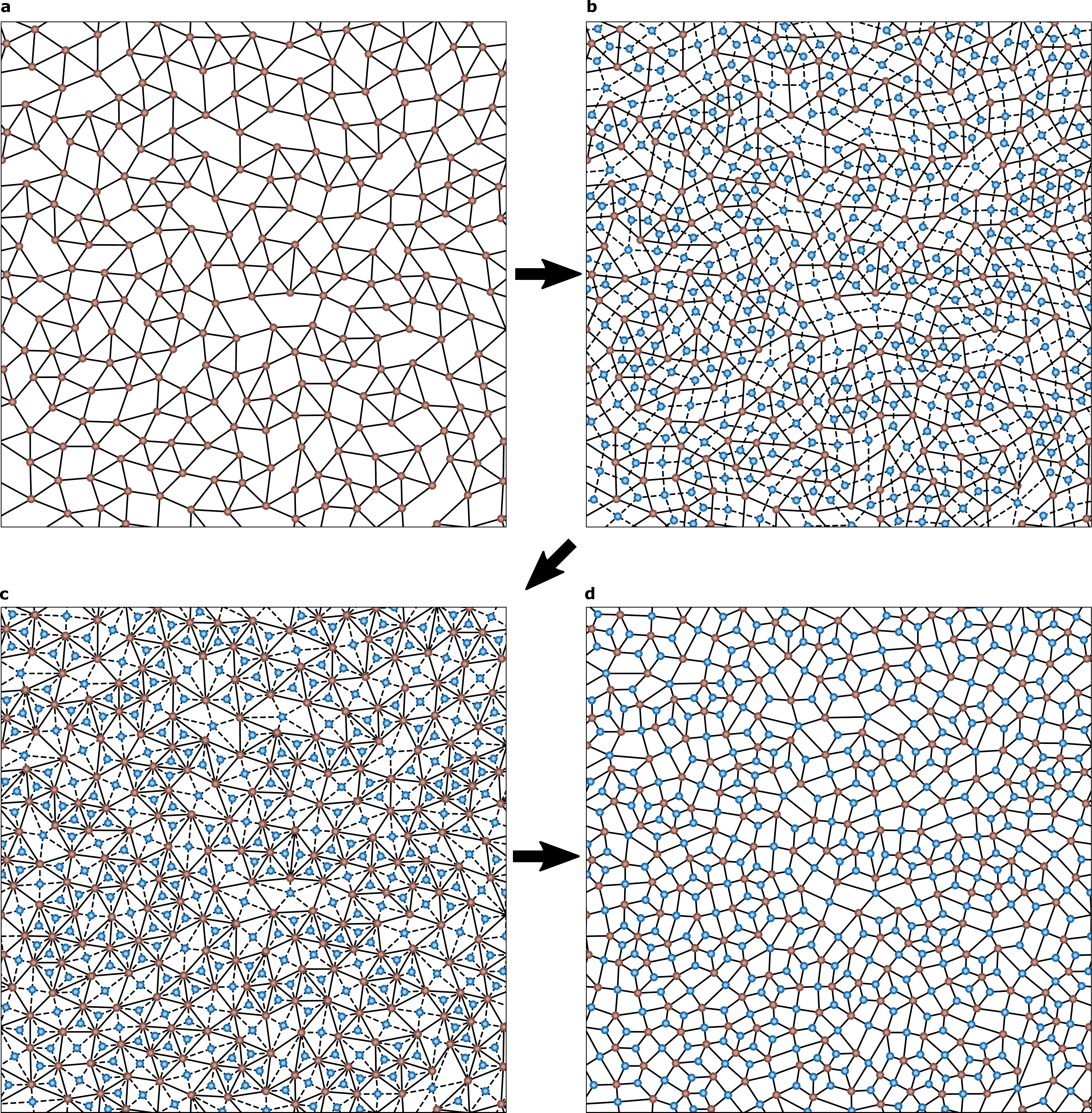}
\caption{\label{fig:bipTrans} \textbf{Planar Bipartite Transformation.} \textbf{a)} Start with a random planar contact network, in this case the coordination is $z=5$. \textbf{b)} Compute the dual graph of the network (blue nodes and dashed lines), place each node of the dual graph inside its corresponding cycle. \textbf{c)} Disconnect each node of the dual graph (blue to blue nodes), and reconnect them to the nodes corresponding to the vertices of its cycle in the original graph (red and blue nodes connected by dashed lines). \textbf{d)} Remove the connections of the original graph (red to red nodes) and keep the connections between the dual graph and the original graph. As both initial graphs are independent sets that don't connect to themselves, the final result is a random planar bipartite graph. }
\end{figure}

One of the main problems in creating a random perfect auxetic material is the construction of a random bipartite planar graph, from which we can construct the polygon network. The graph must be bipartite so that the polygons can counter rotate with respect to each other, and it must be planar so that we can build the polygon network without overlapping them.

We propose two methods to create these graphs. The first is a heuristic pruning algorithm, which takes advantage of the property that a graph with only even cycles will be bipartite. The second is a general transformation that can quickly create bipartite graphs by combining a graph with its dual graph.

\subsection{Pruning Algorithm}

A bipartite graph has only even cycles, where a cycle is the shortest path between a node and itself. We call a cycle even or odd depending on the number of bonds in its path. Here we prune a graph in such a way that all the cycles of the resulting graph are even, transforming the graph into a bipartite graph.

If we have two neighboring cycles that share a bond, and we prune this bond, we will end up with a single cycle. We can think of this operation as an addition of cycles. Where if the starting cycles are both even or odd, the resulting cycle will be even. And if one cycle is even and the other is odd, the resulting cycle is odd.
We can extrapolate this property to a pair of separate odd cycles with only even cycles between them. If we remove a line of bonds between the odd cycles, we will end up with a single even cycle.
Then, if we prune the bonds between all pairs of odd cycles, we will end up with a bipartite graph with only even cycles.

The pruning algorithm we implemented follows some simple steps, we start with a random planar graph with an even number of odd cycles, ideally with a high coordination number, e.g. $z=6$. Next, we place a marker at an odd cycle and use a breadth-first search algorithm on the dual graph, to find the path to its closest odd cycle. We remove the bonds in this path, transforming both odd cycles into a single even cycle. Finally we move the marker to another odd cycle and repeat the procedure until all cycles are even. An example of the initial and final graphs is in Figure \ref{fig:prune}.
While removing bonds we prefer paths that leave each node with at least three bonds, to give the system more stability and to avoid generating big holes. If a node is left with less than two links, we eliminate the node.

This algorithm works well on small graphs, but on bigger ones it may eliminate a huge amount of bonds leaving big holes. To avoid this problem a more sophisticated path optimization algorithm is needed, where the paths between all pairs of odd cycles are calculated beforehand, minimizing the distance of each path and making sure they avoid each other. Once all of the paths are computed, the bond elimination process can be performed, minimizing the size of the holes.
Furthermore, this algorithm does not necessarily work for graphs with periodic boundary conditions. As having only even sided cycles guarantees bipartivity if the graph has free boundary conditions, but it does not if the graph has periodic boundary conditions. 

\subsection{Bipartite Transformation}

A bipartite graph is made out of two independent sets, each one connected to the other but note to itself. Here we connect two independent sets, a graph and its dual graph, transforming both into a single planar bipartite graph.

Given a graph, we first determine its dual graph. Then we connect each node of the graph with each neighboring node in the dual graph, by neighbor node we mean the node in the dual graph that represents a face in contact with the node in the original graph. At last, we eliminate all the starting bonds of the graph and its dual graph, leaving only the new bonds connecting both graphs. The resulting graph will be bipartite, and if the original graph was planar, the resulting graph will be planar too. The further understand this procedure, see Figure \ref{fig:bipTrans}.

This transformation can be performed on any kind of graph and several times in a row creating a graph with more nodes each time. We can reverse the transformation, though we may not know if we obtained the graph or its dual graph when performing the inverse transformation, unless we keep track of at least one node from the starting graph.

\section{Polygon Network Degrees of Freedom}
\label{DoF}

If polygons are taken as rigid structures an extended version of the Maxwell's degrees of freedom counting argument \cite{maxwell1864calculation} can be done. Each polygon has 3 degrees of freedom in two dimension and one hinge (zero length spring) between two polygons suppresses two degrees of freedom. For a network that has $N$ polygons and $N_s$ springs  between polygons, the system is jammed when the number of degrees of freedom $3N$ is less than the number of constraints $2N_s$. Expressing $N_s$ as a function of the coordination (the average number of springs per polygon) $N_s=\frac{zN}{2}$, one gets a critical coordination $z_c=3$. Above critical coordination polygon networks typically guaranteed mechanical stability, due to the absence of trivial floppy modes. Therefore, the existence of an ``auxetic'' floppy mode must be related to a very precise geometrical construction which also implies the appearance of a non trivial self stress state mode following the rank theorem \cite{calladine1978buckminster}.

\section{Numerical Methods}
\label{numerical}

To obtain the data shown in Main Text Figure \ref{fig:frustrations}d and \ref{fig:frustrations}e, we performed numerical simulations of polygon networks. To model these networks, we used the potential energy in equation (\ref{eq:springLam}). Depending on the problem and the boundary conditions we used different methods to compress and test the polygon networks. Regardless of the boundary conditions the Poisson's ratio was calculated using $\nu=-\frac{d\epsilon_x}{d\epsilon_y}=-\frac{dL_x}{dL_y}\frac{L_y}{L_x}$, where $L_x$ and $L_y$ are the approximate dimensions of the system in each axis.

\subsection{Periodic Boundary Conditions}

The polygon networks in Main Text Figure \ref{fig:frustrations}d were set in a periodic boundary box. To compress the material uniaxially, we shrank the box in the vertical direction, and we let the system relax in the horizontal direction by minimizing its energy. At each step we measured the dimensions of the box, $L_x$ and $L_y$.

\subsection{Free Boundary Conditions}

The system in Main Text Figure \ref{fig:frustrations}e has free boundary conditions and an internal floppy mode. To efficiently deform the material, we applied a deformation in the direction of the floppy mode and minimized its energy afterwards. To obtain the floppy mode, we fixed 3 degrees of freedom in the system and found a non-trivial solution for $M\dot{\vec{q}}=0$, where  $M$ is the Hessian and $\dot{\vec{q}}$ is the floppy mode. At each step we approximated the system by a rectangle of dimension $L_x$ and $L_y$.

\section{Finite Element Simulations}
\label{FEM}

\begin{figure}
\centering
\includegraphics[width=\linewidth]{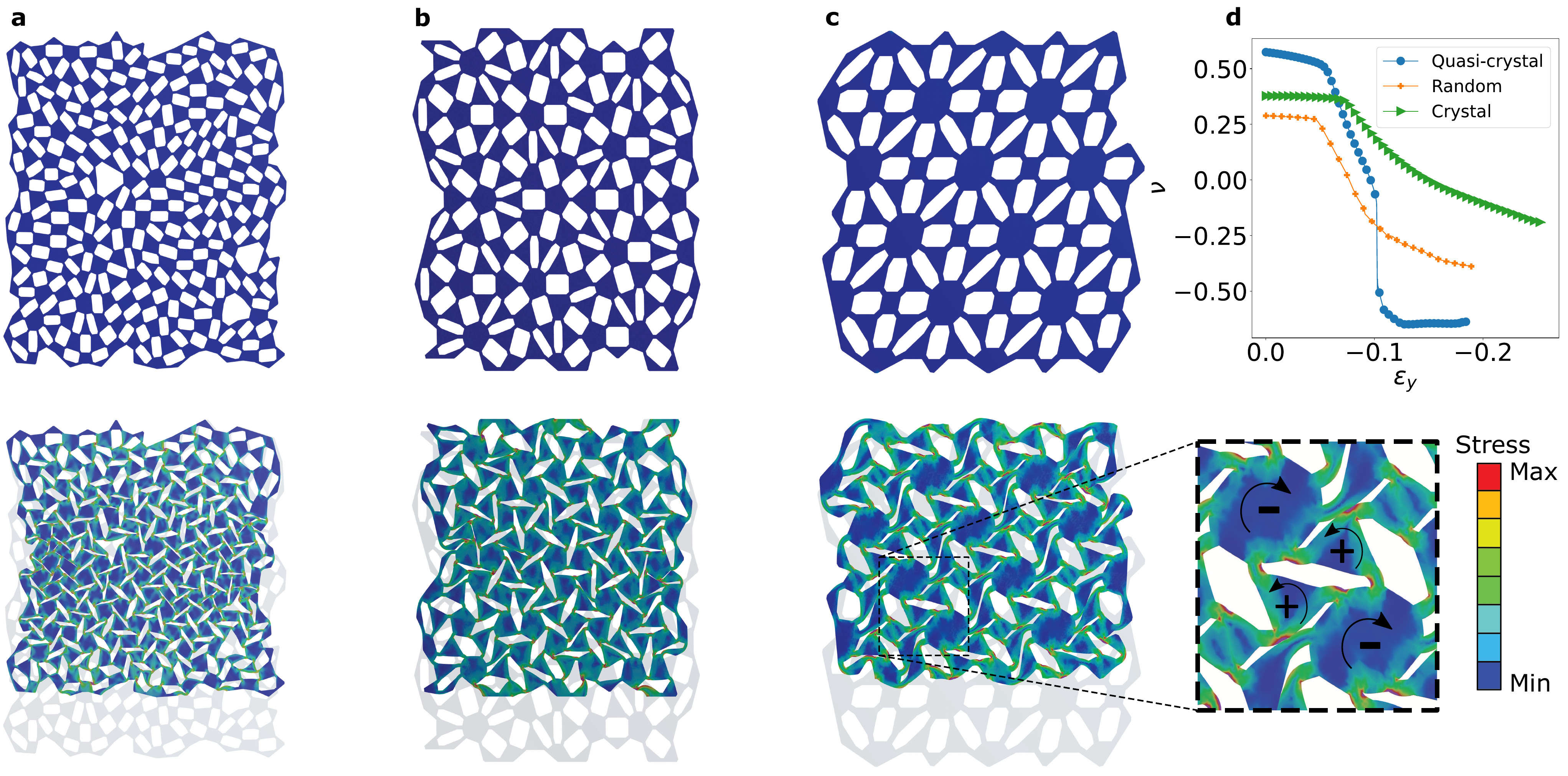}
\caption{\label{fig:BigFem} \textbf{Finite element simulations.} The three systems correspond to the top row a) Random Lattice (isotropic lattice), b) Penrose’s quasicrystal, and c) Exotic Crystal. The structures were simulated using the commercial software Ansys Mechanical \cite{Ansys} with a finite element method. Their uniaxial compression results are depicted in the bottom row where the auxetic behavior is apparent, the color shows the intensity of the stress in the material. Each of the two interconnected lattices that fit the bipartite system rotates in opposite senses, as illustrated in the inset of figure c). Finally, in d), we display the Poisson’s ratios calculated from the simulations.}
\end{figure}

We performed numerical simulations on bigger networks with the same pattern as in Main Text Figure \ref{fig:redes}, to verify that the auxetic behavior is not a particular local phenomena. We kept the principal features in the network's construction as the ratio between average polygon size and the bond thickness. All of the structures show an auxetic behavior and counter-rotation between neighbor polygons as seen in Figure \ref{fig:BigFem}.

To measure the Poisson's ratio, we approximated the whole system as a rectangle with dimensions $L_x$ and $L_y$. Then the strains are $ \Delta \epsilon_x=\frac{L_x-L^{(0)}_x}{L^{(0)}_x}$ and $ \Delta \epsilon_y=\frac{L_y-L^{(0)}_y}{L^{(0)}_y}$\cite{bertoldi2010negative}, where $L_x^0$ and $L_y^0$ are the dimensions of the material at rest.
Finally, we used the engineering strain Poisson's ratio
\begin{equation}
 \nu = -\frac{\Delta \epsilon_x}{\Delta \epsilon_y}.
\end{equation}

\section{Domain Walls}
\label{DomainWalls}

\begin{figure}
\centering
\includegraphics[width=0.5\linewidth]{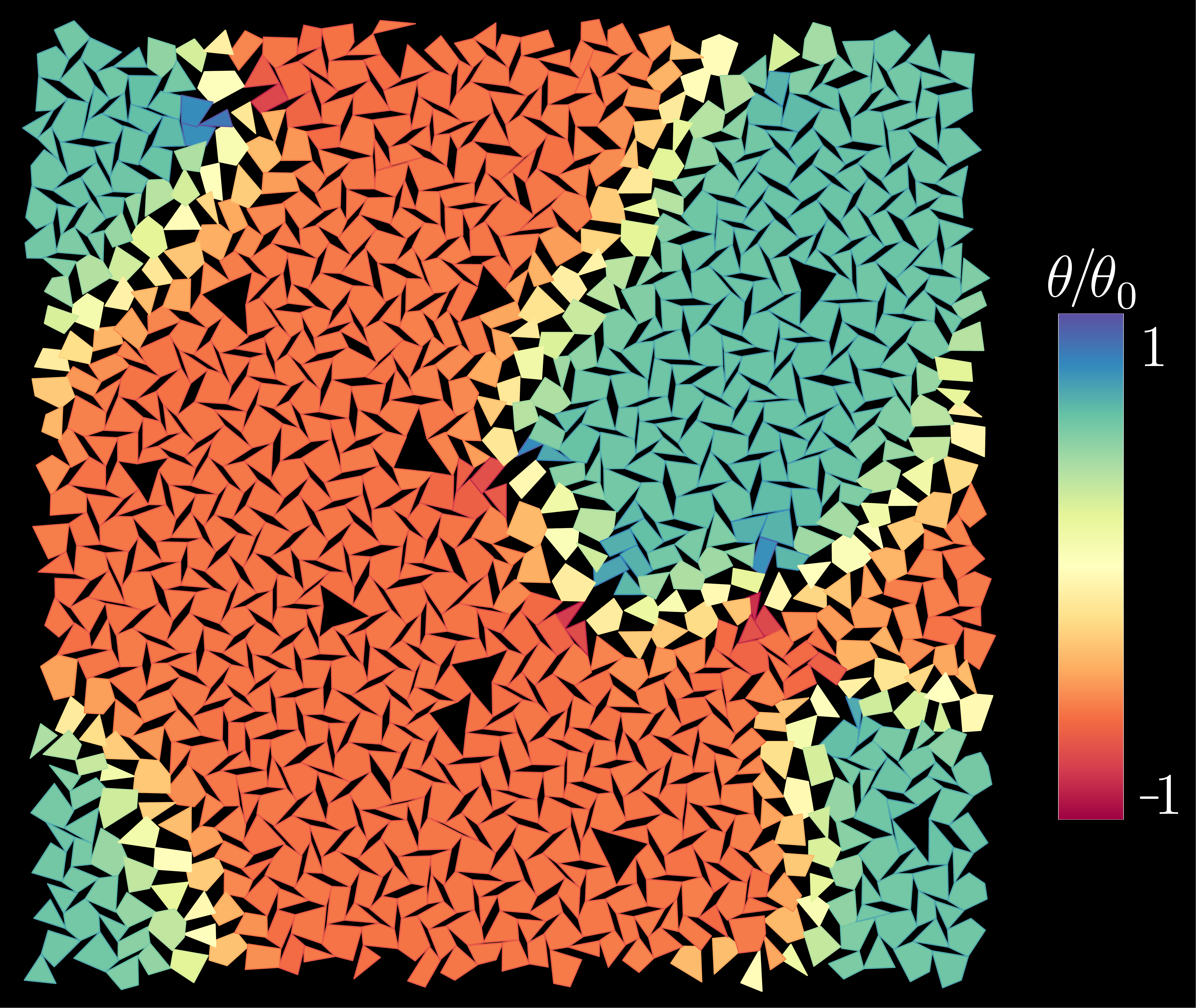}
\caption{\label{fig:domWall} \textbf{Domain wall in a random auxetic.} The system has periodic boundary conditions and it is compressed with $\delta \lambda=0.2$. The color of each polygon represents their normalized angle of rotation. There are two domains one in red and the other in blue, both separated by a yellow domain wall.}
\end{figure}

\begin{figure}
\centering
\includegraphics[width=0.5\linewidth]{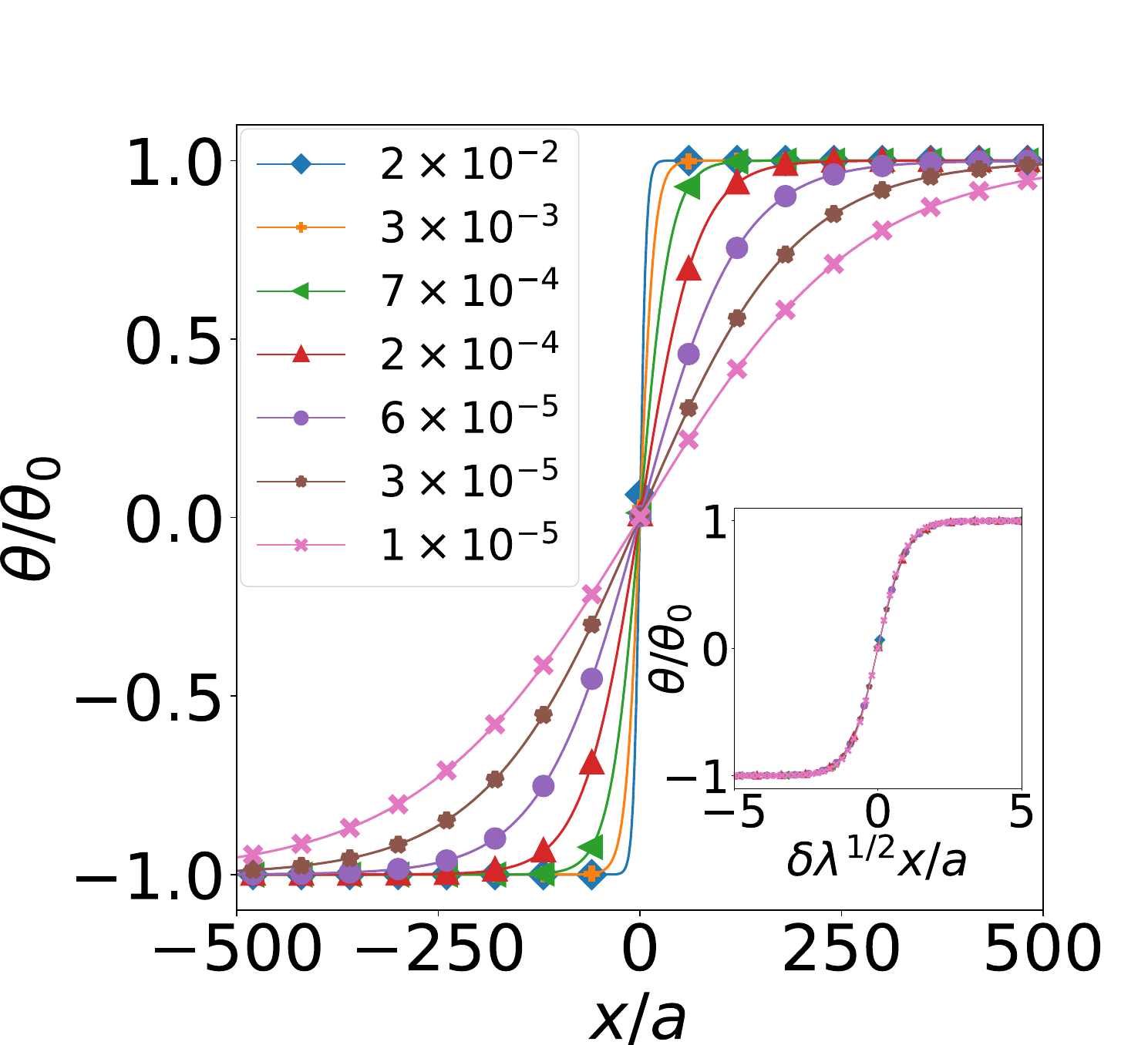}
\caption{\label{fig:deltaVlambda}  \textbf{Length scale of the domain wall width.} The normalized rotational angle of the polygons as a function of distance, revealing the existence of a domain wall in the middle. Each curve represents a polygon network with different levels of compression, measured by $\delta \lambda$. In the inset the x-axis has been rescaled by $\delta \lambda^{-1/2}$, all the curves collapse into a single one showing that it's the correct scaling. The average distance from a polygon's position to its vertices is $a=1/2$. The normalization coefficient $\theta_0$ corresponds to the maximum rotation angle of the system. We used a periodic polygon network as the one in Main Text Figure \ref{fig:frustrations}c with a vertical domain wall.}
\end{figure}

We can see that Main Text equation (\ref{eq:ising}) is analogous to that of an antiferromagnetic spin system. In such case, $\theta$ represents the spin direction, $J_{ij}$ is a symmetric coupling constant and the sum over neighbors $H_i=\sum_n H_{in}$ is a magnetic field as a function of space.

Therefore, we can find magnetic related phenomena, like domain walls, in polygon networks. These appear between two stable solutions of the system which differ in the turning direction of the polygons, as seen in Figure \ref{fig:domWall}. These domain walls have been previously studied in auxetics\cite{courentinDomainWall,BertoldiDomainWall}.

The polygon angle defines the state of the polygon network, when $0<\lambda\leq 1$ the angle is zero and when $1<\lambda<\big|\frac{1+C}{1-C}\big|$ the angle is given by Main Text equation (\ref{eq:minima1}) and (\ref{eq:minima2}). Then $\theta$ can be used as the order parameter of this system which is controlled by the external parameter $\delta \lambda=\lambda-1$. We can now build a simple Ginzburg-Landau energy for the system and make the usual analysis for the domain walls in it \cite{kardar_2007}.

\begin{equation}
\label{eq:GLenergy}
H=\int dx \left( \frac{t}{2}\theta(x)^2+u\theta(x)^4+\frac{K}{2}\left(\nabla\theta(x)\right)^2 \right)
\end{equation}

Here $t$, $u$ and $K$ are analytical functions of $\delta \lambda$. To keep the system stable $K$ and $u$ are positive constants close to the critical point, and $t=-t_0\delta \lambda+O(\delta \lambda)^2$.

The stable solution for an homogeneous system is given by:

\begin{eqnarray}
\bar{\theta}= \sqrt{\frac{t_0}{4u}}\delta\lambda^{1/2}.
\end{eqnarray}

To determine the length scale of a domain wall we search for the minimum energy in a system with $\theta(\infty)=\bar\theta$,  $\theta(-\infty)=-\bar\theta$. The differential equation for such system is:

\begin{eqnarray}
K\frac{d^2 \theta}{dx^2}=t\theta+4u\theta^3,
\end{eqnarray}

whose well known solution is:

\begin{equation}
\theta(x)=\bar\theta \tanh\left(\frac{x}{\Delta}\right).
\end{equation}

Where the domain wall width is defined as:

\begin{equation}
\label{eq:domainwall}
\Delta=\sqrt{\frac{2K}{t_0}}\delta \lambda^{-1/2}.
\end{equation}

Thus we see that the domain wall width scales like $\Delta \sim \delta \lambda^{-1/2}$. To check this, we performed numerical simulations where we minimized the energy of a polygon network with periodic boundary conditions until it had a couple of stable domain walls. To simplify the problem we used flat domain walls, the results can be seen in Figure \ref{fig:deltaVlambda}. 

In Figure \ref{fig:deltaVlambda} we consider that the polygon angle as a function of position is $\theta(x)=\theta_0 f(x)$ with $|f(x)|\leq 1$,  the expansion of this function around the origin is $f(x)=mx+O(x^3)$, as $f$ is an even function. Now through simple trigonometry we can relate the slope at the origin $m$ to the domain wall length, approximately $\Delta=2/m$, and from the rescaled inset in Figure \ref{fig:deltaVlambda} we can determine that $m\sim\delta \lambda^{1/2}$, therefore we obtained the same result as predicted where $\Delta\sim\delta \lambda^{-1/2}$.

\section{Perfect Auxetic Detailed Derivation}
\label{PerfectAuxeticDerivation}

In this section we will find the zero energy auxetic mode of a perfect auxetic polygon network, this is a polygon network that follows 3 requirements.

1.- The underlying graph of the connected polygons must be bipartite, this means that we can split the graph into two sets, where each set connects to the other but not to itself, we will call each set $A$ and $B$.

2.- At rest and without prestress the system must have at least one configuration where every pair of neighboring polygons positions ($\vec{x}_i$, $\vec{x}_j$) are collinear with the vertex between them.
 
3.- If we set a point in space representing each polygon position $\vec{x}_i$, not necessarily the centroid of each polygon, the distance from this position to each vertex will be $a_{ij}$ in polygons of set $A$, and $b_{ji}$ in polygons of set $B$, the first index indicates the origin polygon and the second index represents the neighboring polygon, finally the ratio of this distances between neighboring polygons must remain constant, so $C=\frac{b_{ji}}{a_{ij}}$ is constant.

The following demonstration applies to undercontrained and overconstrained systems, though the later are of higher interest as this zero mode is non-trivial in them. 

To show the existence of this zero mode, we will expand its potential energy and find its minimum under the assumption that the distance between polygons remains constant and that all polygons in each set rotate equally. Finally we will show that the minimum of energy is zero for any value of the expansion coefficient $\lambda$.

Using restriction 1, as the system is bipartite, we can write its potential energy as an interaction of each set $A$ and $B$.

\begin{eqnarray}
\label{eq:PerfectAuxPot}
V=\frac{k}{2}\sum_{<ij>}\left( (\vec{x}_i+\lambda\vec{a}_{ij})  -(\vec{x}_j+\lambda\vec{b}_{ji})  \right)^2
\end{eqnarray}

Here $\vec{x}_i$ is the position of the polygon $i$, $\lambda$ is an homogeneous expansion coefficient, $\vec{a}_{ij}$ and $\vec{b}_{ji}$ are the vectors of the set $A$ and $B$ respectively that point from the position of a polygon to its vertex associated with the pair $<ij>$. They are defined as

$$\vec{a}_{ij}=a_{ij}
\begin{pmatrix}
           \cos (\theta_i+\alpha_{ij})
           \sin (\theta_i+\alpha_{ij})
         \end{pmatrix}$$
         
         and

$$\vec{b}_{ji}=b_{ji}
\begin{pmatrix}
           \cos (-\theta_j-\beta_{ji}) 
           \sin (-\theta_j-\beta_{ji})
         \end{pmatrix}.         
         $$

Note that we have set-up the variables $\theta_i$ and $\theta_j$ such that they naturally counter rotate each other. We will use the index $i$ for the polygons in the set $A$ and $j$ for the polygons in set $B$.

Expanding equation (\ref{eq:PerfectAuxPot}),
\begin{equation}
\label{eq:PerfectAuxPotExpanded}
V=\frac{k}{2}\sum_{<ij>}  (\vec{x}_i-\vec{x}_j)^2+\lambda^2(\vec{a}_{ij}-\vec{b}_{ji})^2+2\lambda(\vec{x}_i-\vec{x}_j)(\vec{a}_{ij}-\vec{b}_{ji}).
\end{equation}

Now will review each term of the expansion of equation (\ref{eq:PerfectAuxPotExpanded}).

Taking into account condition 2, if the rest state of the system is at $\lambda=1$ and $\theta_i=0$, then as the polygon's positions are collinear with the vertex between them, the angles pointing to this vertex must add to half of a full rotation $|\alpha_{ij}+\beta_{ji}|=\pi$. Then,

\begin{equation}
\label{eq:PerexpT1}
(\vec{a}_{ij}-\vec{b}_{ji})^2=a_{ij}^2+b_{ji}^2+2a_{ij}b_{ji}\cos(\theta_i+\theta_j).
\end{equation}

Moreover if we assume that the distance between polygon position's remains constant as $\lambda$ is increased, we can express the distance as
         
$$\vec{x}_{i}-\vec{x}_{j}=-(a_{ij}+b_{ji})
\begin{pmatrix}
           \cos (\alpha_{ij})
           \sin (\alpha_{ij})
         \end{pmatrix}.$$

Thus we can determine the other two terms of equation (\ref{eq:PerfectAuxPotExpanded}).

\begin{equation}
\label{eq:PerexpT2}
(\vec{x}_i-\vec{x}_j)^2=(a_{ij}+b_{ji})^2
\end{equation}

and

\begin{equation}
\label{eq:PerexpT3}
(\vec{x}_i-\vec{x}_j)(\vec{a}_{ij}-\vec{b}_{ji})=-(a_{ij}+b_{ji})(a_{ij}\cos(\theta_i)+b_{ji}\cos(\theta_j)).
\end{equation}

Using equation (\ref{eq:PerexpT1}), (\ref{eq:PerexpT2}) and (\ref{eq:PerexpT3}) to expand equation (\ref{eq:PerfectAuxPotExpanded}), we arrive to

\begin{equation}
\label{eq:PerfectAuxPotFinal}
V=\frac{k}{2}\sum_{<ij>} (a_{ij}+b_{ji})^2+\lambda^2(a_{ij}^2+b_{ji}^2+2a_{ij}b_{ji}\cos(\theta_i+\theta_j))-2\lambda(a_{ij}+b_{ji})(a_{ij}\cos(\theta_i)+b_{ji}\cos(\theta_j)).
\end{equation}

We can rewrite this as:

\begin{equation}
\label{eq:PerfectAuxPotFinal2}
V=V_0+\sum_{<ij>}J_{ij}\cos(\theta_i+\theta_j)-H^A_{ij}\cos(\theta_i)-H^B_{ji}\cos(\theta_j).
\end{equation}

Here $V_0=\frac{k}{2}\sum_{<ij>} (a_{ij}+b_{ji})^2+\lambda^2(a_{ij}^2+b_{ji}^2)$, $J_{ij}=k \lambda^2 a_{ij}b_{ji}$, $H^A_{ij}=k\lambda (a_{ij}+b_{ji})a_{ij}$ and $H^B_{ji}=k\lambda (b_{ji}+a_{ij})b_{ji}$.

We can search for a minimum in equation (\ref{eq:PerfectAuxPotFinal}), if we assume that all polygons in each set rotate equally. Then $\theta_i=\theta_A$ and $\theta_j=\theta_B$. And if we use the requirement 3, setting the ratio $C=\frac{b_{ji}}{a_{ij}}$ as a constant, the energy becomes

\begin{equation}
\label{eq:PerfectAuxPotFinalC}
V=( (C+1)^2+\lambda^2(C^2+1+2C\cos(\theta_A+\theta_B)) -2\lambda(C+1)(\cos(\theta_A)+C\cos(\theta_B)))\frac{k}{2}\sum_{<ij>}a_{ij}^2.
\end{equation}

Deriving by $\theta_A$ and $\theta_B$, we have

\begin{equation}
\label{eq:PerfectAuxSol1}
-2\lambda^2C\sin(\theta_A^0+\theta_B^0)+2\lambda(1+C)\sin(\theta_A^0)=0
\end{equation}

and

\begin{equation}
\label{eq:PerfectAuxSol2}
-2\lambda^2C\sin(\theta_A^0+\theta_B^0)+2\lambda(1+C)C\sin(\theta_B^0)=0.
\end{equation}

From equation (\ref{eq:PerfectAuxSol1}) and (\ref{eq:PerfectAuxSol2}), we see that $\sin(\theta_A^0)=C\sin(\theta_B^0)$, plugging back this into equation (\ref{eq:PerfectAuxSol1}) we find that $\sin(\theta_A^0)=0$, which is the trivial minimum $\theta_A^0=\theta_B^0=0$ for $0<\lambda\leq 1$. Leaving equation (\ref{eq:PerfectAuxSol1}) like

\begin{equation}
C\cos(\theta_B^0)+\cos(\theta_A^0)=\frac{(1+C)}{\lambda}.
\end{equation}

We simplify this equation using $\sin(\theta_A^0)=C\sin(\theta_B^0)$.

\begin{equation}
\label{eq:PerfectAuxMin1}
\cos(\theta_A^0)=\frac{1+C+\lambda^2(1-C)}{2\lambda}
\end{equation}

\begin{equation}
\label{eq:PerfectAuxMin2}
\cos(\theta_B^0)=\frac{1+C+\lambda^2(C-1)}{2\lambda C}
\end{equation}

This solution is a minimum of energy for $1<\lambda<\big|\frac{1+C}{1-C}\big|$.
We will now replace it into the potential energy in equation (\ref{eq:PerfectAuxPotFinalC}), selecting the solutions where both $\theta_A$ and $\theta_B$ have the same sign and using that $\sin(\theta_A^0)=C\sin(\theta_B^0)$, we can write equation (\ref{eq:PerfectAuxPotFinalC}) as,
 
\begin{equation}
V=( (C+1)^2+\lambda^2(1-C^2+2C\cos(\theta_B^0)(\cos(\theta_A^0)+C\cos(\theta_B^0)))-2\lambda(C+1)(\cos(\theta_A^0)+C\cos(\theta_B^0)))\frac{k}{2}\sum_{<ij>}a_{ji}.
\end{equation}

Now using that $\cos(\theta_A^0)+C\cos(\theta_B^0)=\frac{1+C}{\lambda}$,
 
\begin{equation}
V=( (C+1)^2+(C+1)^2-2(C+1)^2)\frac{k}{2}\sum_{<ij>}a_{ji}^2.
\end{equation}

Which is clearly zero, therefore we find that $V(\theta_i=\theta^0_A,\theta_j=\theta^0_B)=0$ for $1<\lambda<\big|\frac{1+C}{1-C}\big|$. This proves that the system has a zero energy mode that expands the polygon network isotropically, i.e. it's a perfect auxetic.

\section{Zero Bulk Modulus}
\label{bulk}
If an elastic material has a zero bulk modulus, it means that it has a floppy mode that allows the material to freely deform isotropically, i.e. it has a Poisson's ratio $\nu=-1$.

Here we will show that a polygon network that follows our three rules has a bulk modulus $K=0$ and is therefore a perfect auxetic. Furthermore, its shear modulus will remain finite, as the system is overconstrained.

Starting from the potential energy, we can write it as a function of a small isotropic expansion of each polygon $\delta \lambda$.

\begin{equation}
V=\sum_{\langle i~j\rangle} \frac{k}{2} {\vec{r}_{ij}}^{~2},
\end{equation}

with $\vec{r}_{ij}= \vec{x}_i-\vec{x}_j +  (1+\delta \lambda)(\vec{a}_{ij}-\vec{b}_{ji})$
the bulk modulus will be given by the coefficient of a linear expansion of the energy as a function of $\delta \lambda$.

\begin{equation}
\Omega K=\frac{d^2V}{d\delta \lambda^2}|_{\delta \lambda=0},
\end{equation}

where $\Omega$ is the area of the system. If we expand this last equation we obtain,

\begin{equation}
\label{eq:Ksimple}
\Omega K=k\sum_{\langle i~j\rangle}\left(\left(\frac{d\vec{r}_{ij}}{d \delta \lambda}\right)^{2}+\left(\frac{d^2\vec{r}_{ij}}{d {\delta \lambda}^2}\right)\cdot \vec{r}_{ij} \right)|_{\delta \lambda=0}.
\end{equation}

If we assume that the system is unstressed, then the length of each spring must be zero $\vec{r}_{ij}=0$, and from equation (\ref{eq:Ksimple}) if $K=0$ we see that $\frac{d\vec{r}_{ij}}{d \delta \lambda}=0$, which means that the spring length must be zero as a function of $\delta \lambda$. This is quite reasonable as zero bulk modulus implies that there is a zero mode in the system which doesn't deform the springs.
Expanding $\frac{d\vec{r}_{ij}}{d \delta \lambda}=0$ we find that,

\begin{equation}
 \dot{\theta}_i\vec{a}'_{ij} -\dot{\theta}_j \vec{b}'_{ji}+\dot{\vec{x}}_i-\dot{\vec{x}}_j=-\vec{a}_{ij}+\vec{b}_{ji}.
\end{equation}

Where dots are derivatives by $\delta \lambda$ and primes are derivatives by the corresponding angle $\theta$. If we now assume that the polygons only rotate and have no displacements while expanding, i.e. $\dot{\vec{x}}=0$. And if our network is bipartite, with each independent set rotating with the same angle ${\theta}_i= \theta_{A}$ and ${\theta}_j=\theta_{B}$, our equation becomes

\begin{equation}
 \dot{\theta}_A\vec{a}'_{ij} -\dot{\theta}_B \vec{b}'_{ji}=-\vec{a}_{ij}+\vec{b}_{ji}.
\end{equation}

Taking the projection of this equation in the direction of $\vec{a}'_{ij}$ and $\vec{b}'_{ji}$ we obtain a vectorial expression for the angular velocity.

\begin{equation}
 \dot{\theta}_{A}=\frac{\vec{a}_{ij}\cdot \vec{b}'_{ji}   \vec{a}'_{ij}\cdot \vec{b}'_{ji}}{\vec{a}'^2_{ij}\vec{b}'^2_{ji}-\left( \vec{a}'_{ij}\cdot\vec{b}'_{ji} \right)^2}
\end{equation}

\begin{equation}
 \dot{\theta}_{B}=\frac{\vec{b}_{ji}\cdot \vec{a}'_{ij}   \vec{a}'_{ij}\cdot \vec{b}'_{ji}}{\vec{a}'^2_{ij}\vec{b}'^2_{ji}-\left( \vec{a}'_{ij}\cdot\vec{b}'_{ji} \right)^2}
\end{equation}

Following the three rules for perfect auxetics in Appendix \ref{PerfectAuxeticDerivation}. We can use the following representation of the vectors $\vec{a}_{ij}$ and $\vec{b}_{ji}$ to easily operate them.

$$\vec{a}_{ij}=a_{ij}
\begin{pmatrix}
           \cos (\theta_i+\alpha_{ij})
           \sin (\theta_i+\alpha_{ij})
         \end{pmatrix}$$

$$\vec{b}_{ji}=b_{ji}
\begin{pmatrix}
           \cos (-\theta_j-\beta_{ji})
           \sin (-\theta_j-\beta_{ji})
         \end{pmatrix}         
         $$
With $b_{ji}/a_{ij}=C$ and $|\alpha_{ij}+\beta_{ji}|=\pi$.

\begin{equation}
\dot{\theta}_{A}= \frac{C+\cos(\theta_A+\theta_B)}{\sin(\theta_A+\theta_B)}
\end{equation}
\begin{equation}
\dot{\theta}_{B}= \frac{1+C\cos(\theta_A+\theta_B)}{C\sin(\theta_A+\theta_B)}
\end{equation}

As $\dot{\theta}$ only depends on general properties of the whole system, it is a valid solution for small deformations in a system with $K=0$. Therefore, a system that follows the stipulated rules, will be a perfect auxetic.

\end{document}